\input{aipcheck}

\documentclass[
    ,final            
  ]
  {aipproc}

\layoutstyle{8x11double}

\usepackage{graphicx}
\usepackage{epsfig}
\usepackage{latexsym}
\usepackage{amsmath}
\usepackage{amsfonts}
\usepackage{amsxtra}
\usepackage{axodraw}

\def\krto{ {\,\,\lower .8ex\hbox {$\longrightarrow \atop k \rightarrow 0$}\,\,}}

\def\bea{\begin{eqnarray} }
\def\beq{\begin{eqnarray} }

\def\eea{\end{eqnarray}}
\def\eeq{\end{eqnarray}}
\def\eq#1{eq.~(\ref{#1})}

\begin{document}

\title{The scaling infrared DSE solution as a critical end-point for the family of decoupling ones}

\classification{\texttt{12.35.Aw, 12.38.Lg, 12.38.Gc}}
\keywords{Dyson-Schwinger equations, Infrared QCD}

\author{J. Rodr\'{\i}guez-Quintero}{
  address={Dpto. F\'isica Aplicada, Fac. Ciencias Experimentales; 
Universidad de Huelva, 21071 Huelva; Spain}
}



\begin{abstract}

Both regular (the zero-momentum 
ghost dressing function not diverging), also named decoupling, and 
critical (diverging), also named scaling, Yang-Mills propagators solutions 
can be obtained by analyzing the low-momentum behaviour of the ghost propagator Dyson-Schwinger 
equation (DSE) in Landau gauge. The asymptotic expression obtained for the 
regular or decoupling ghost dressing function up to the order ${\cal O}(q^2)$ fits pretty well 
the low-momentum ghost propagator obtained through the numerical integration of 
the coupled gluon and ghost DSE in the PT-BFM scheme. Furthermore, when the size of the 
coupling renormalized at some scale approaches some critical value, the PT-BFM results
seems to tend to the the scaling solution as a limiting case. 

\end{abstract}

\maketitle


\section{Introduction}

The low-momentum behaviour of the Yang-Mills propagators derived either from 
the tower of Dyson-Schwinger equations (DSE) or from Lattice simulations in 
Landau gauge has been a very interesting and hot topic for the 
last few years. It seems by now well established that, if we assume 
in the vanishing momentum limit a ghost dressing function behaving as 
$F(q^2) \sim (q^2)^{\alpha_F}$ and a gluon propagator as 
$\Delta(q^2) \sim (q^2)^{\alpha_G-1}$ (or, by following a notation commonly used,
a gluon dressing function as $G(q^2)= q^2 \Delta(q^2) \sim (q^2)^{\alpha_G}$), 
two classes of solutions may emerge (see, for instance, the discussion 
of refs.~\cite{Boucaud:2008ji,Boucaud:2008ky}) from the DSE:
(i) those, dubbed {\it ``decoupling''}, where $\alpha_F=0$ and the suppression of 
the ghost contribution to the gluon propagator DSE results in a massive gluon 
propagator (see \cite{Aguilar:2006gr,Aguilar:2008xm} and references therein); 
and (ii) those, dubbed {\it ``scaling''}, where $\alpha_F \neq 0$ 
and the low-momentum behaviour of both gluon and ghost propagators 
are related by the coupled system of DSE through the condition $2 \alpha_F+\alpha_G = 0$ 
implying that $F^2(q^2)G(q^2)$ goes to a non-vanishing constant when $q^2 \to 0$ 
(see \cite{Alkofer:2000wg,Fischer:2008uz} and references therein). 

Lattice QCD results appear to support only the massive gluon ($\alpha_G=1$) or scaling solutions  
(see~\cite{Cucchieri:2007md} 
and references therein), and also pinching technique results (see, for instance, 
\cite{Cornwall,Binosi:2002ft} and references therein), 
refined Gribov-Zwanziger 
formalism (see~\cite{Dudal:2007cw}) or other approaches like the infrared mapping of $\lambda \phi^4$ and Yang-Mills 
theories in ref.~\cite{Frasca:2007uz} or the massive extension of the Fadeev-Popov 
action in ref.~\cite{Tissier:2010ts} appear to point to.
In the present note, we briefly review the work of ref.~\cite{RodriguezQuintero:2010xx}, 
which extended the previous studies of refs.~\cite{Boucaud:2008ji,Boucaud:2008ky,Boucaud:2010gr}, 
by the analysis of the results obtained by solving the coupled system of Landau gauge ghost and gluon propagators DSE 
within the framework of the pinching technique in the background field method~\cite{Binosi:2002ft} (PT-BFM) 

\section{The two kinds of solutions of the ghost propagator Dyson-Schwinger equation}\label{revisiting}
\label{twosol}

As was explained in detail in refs.~\cite{Boucaud:2008ky,Boucaud:2010gr,RodriguezQuintero:2010xx}, the low-momentum behavior 
for the Landau gauge ghost dressing function can be inferred from the analysis
of the Dyson-Schwinger equation for the ghost
propagator (GPDSE). That analysis is performed on a very general ground: one applies the MOM renormalization 
prescription,
$F_R(\mu^2) =  \mu^2 \Delta_R(\mu^2) = 1$, 
where $\mu^2$ is the subtraction point, chooses for the ghost-gluon vertex, 
\beq
\widetilde{\Gamma}_\nu^{abc}(-q,k;q-k) &=& 
i g_0 f^{abc} \left( q_\nu H_1(q,k) \right. \nonumber \\
&+& 
\left. (q-k)_\nu H_2(q,k) \right) 
\label{DefH12}
\eeq
to apply this MOM prescription in Taylor kinematics 
({\it i.e.} with a vanishing incoming ghost momentum) 
and assumes the non-renormalizable bare ghost-gluon form factor, $H_1(q,k)=H_1$, 
to be constant in the low-momentum regime for the incoming ghost.
Then, the low momentum-behaviour of the ghost dressing function is 
supposed to be well described by 
\beq\label{dress}
F_R(q^2) &=& A(\mu^2) \left( \frac{q^2}{M^2} \right)^{\alpha_F} \left( 1 + \cdots 
\rule[0cm]{0cm}{0.6cm} \right) \ ,
\eeq
and that of the gluon propagator by
\beq\label{gluonprop}
\Delta_R(q^2) &=& \frac{B(\mu^2)}{q^2 + M^2} \ 
\simeq \frac{B(\mu^2)}{M^2} \left( 1 - \frac{q^2}{M^2} + \cdots \right) \ ,
\eeq
and this, after solving asymptotically the GPDSE, finally left us with:
\beq \label{solsFsneq}
F_R(q^2) \simeq 
\left(
\frac {10 \pi^2}{N_C H_1 g_R(\mu^2) B(\mu^2)} 
\right)^{1/2}
\ \left(\frac {M^2} {q^2} \right)^{1/2} \ ,
\eeq
if $\alpha_F \neq 0$; and 
\beq\label{solsFseq}
F_R(q^2) &\simeq&
F_R(0) \left( 1 +   
\frac{N_C H_1}{16 \pi} \ \overline{\alpha}_T(0) \ 
\frac{q^2}{M^2} \left[ \ln{\frac{q^2}{M^2}} - \frac {11} 6 \right] \right.
\nonumber \\
&+& \left. {\cal O}\left(\frac{q^4}{M^4} \right) \right)
\eeq
if $\alpha_F = 0$, 
where 
\beq\label{coefC2}
\overline{\alpha}_T(0) =  M^2 \frac{g^2_R(\mu^2)}{4 \pi} 
F_R^2(0) \Delta_R(0) \ .
\eeq
It should be understood that the subtraction momentum for all the renormalization quantities is $\mu^2$. 
The case $\alpha_F \neq 0$ leads to the so-called scaling solution, where the low-momentum behavior of 
the massive gluon propagator forces the ghost dressing function to diverge at low-momentum 
through the scaling condition: $2 \alpha_F + \alpha_G=0$ ($\alpha_G=1$ is the power 
exponent when dealing with a massive gluon propagator). As this scaling condition is verified, 
the perturbative strong coupling defined in this Taylor scheme~\cite{Boucaud:2008gn}, 
$\alpha_T=g_T^2/(4\pi)$, has to reach a constant at zero-momentum,
\beq
\alpha_T(0) &=& \frac{g^2(\mu^2)}{4 \pi} \lim_{q^2\to 0} q^2 \Delta(q^2) F^2(q^2) \nonumber \\
&=& \frac{5 \pi}{2 N_C H_1} \ ,
\eeq
as can be obtained from Eqs.(\ref{gluonprop},\ref{solsFseq}). 
The case $\alpha_F=0$ corresponds to the so-called decoupling solution, 
where the zero-momentum ghost dressing function reaches a non-zero finite value 
and \eq{solsFseq} provides us with the first asymptotic corrections to this 
leading constant. This subleading correction is controlled by the zero-momentum value of 
the coupling defined in \eq{coefC2}, which is an extension of the non-perturbative effective 
charge definition from the gluon propagator~\cite{Aguilar:2008fh} to the Taylor 
ghost-gluon coupling~\cite{Aguilar:2009nf}. 

\subsection{The ``critical'' limit}

\begin{figure}
\includegraphics[height=.22\textheight]{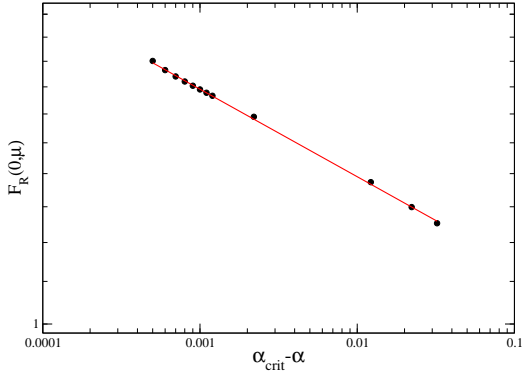} 
\caption{\small Log-log plot of the zero-momentum values of the ghost dressing function, obtained 
by the numerical integration of the coupled DSE system in the PT-BFM scheme, in terms of 
$\alpha_{\rm crit}-\alpha$. $\alpha=\alpha(\mu=10 \mbox{\rm GeV})$, 
the value of the coupling at the renormalization momentum, is an initial condition for 
the integration; while $\alpha_{\rm crit}$ is fixed to be 0.1822, as explained in the text, 
by requiring the best linear correlation.}
\label{fig:ghost0s}
\end{figure}

In ref.~\cite{RodriguezQuintero:2010xx}, the solutions of the coupled DSE system 
in the PT-BFM scheme (with $H_1=1$ for the ghost-gluon vertex), 
numerically integrated for many values of the coupling at 
the renormalization point $\mu^2$ as a boundary condition, 
were studied. They were shown to behave asymptotically as 
\eq{solsFsneq} predicts for the decoupling DSE solutions. 
There also appeared to be a {\it critical} value of the coupling, 
$\alpha_{\rm crit}=\alpha(\mu^2)\simeq 0.182$ with $\mu=10\mbox{\rm ~Gev}$, 
above which the coupled DSE system does not converge any longer to a solution.
As a matter of fact, we know from refs.\cite{Boucaud:2008ky,RodriguezQuintero:2010xx} that
the scaling solution implies for the coupling
\beq\label{ap:crit}
\alpha_{\rm crit} \ = \ \frac{g_R^2(\mu^2)}{4 \pi} 
\simeq
\frac{5 \pi}{2 N_C A^2(\mu^2) B(\mu^2) } \ ,
\eeq
where $B(\mu^2)$ and $A(\mu^2)$ defined by Eqs.~(\ref{gluonprop},\ref{solsFsneq}).
This is also shown in ref.~\cite{Boucaud:2008ji}, where only the ghost propagator 
DSE with the kernel for the gluon loop integral is obtained from gluon propagator lattice 
data. In the analysis of ref.~\cite{Boucaud:2008ji}, a ghost dressing 
function solution diverging at vanishing momentum appears to exist and 
verifies eqs.~(\ref{solsFsneq},\ref{ap:crit}), while regular or decoupling solutions
 exist for any $\alpha < \alpha_{\rm crit}$. In ref.~\cite{RodriguezQuintero:2010xx}, 
 a more complete analysis is performed: first by studying the solutions for many different values 
 of the coupling, $\alpha=\alpha(\mu^2)$, of a coupled DSE system; and then by showing that 
 the ghost dressing function at vanishing momentum, $F(0,\mu^2)$, 
is described by the following power behaviour,
\beq
F(0) \ \sim \ (\alpha_{\rm crit} - \alpha(\mu^2))^{- \kappa(\mu^2)} \ ,
\eeq
where $\kappa(\mu^2)$ is a critical exponent (depending presummably on the 
renormalization point, $\mu^2$), supposed to be positive and to govern the transition 
from decoupling ($\alpha < \alpha_{\rm crit}$) to the scaling ($\alpha = \alpha_{\rm crit}$) 
solutions; and where we let $\alpha_{\rm crit}$ be a free parameter to be fitted 
by requiring the best linear correlation for $\log[F(0)]$ in terms of 
$\log[\alpha_{\rm crit}-\alpha]$. In doing so, the best correlation 
coefficient is 0.9997 for $\alpha_{\rm crit}=0.1822$, which is pretty close to 
the critical value of the coupling above which the coupled DSE system does not converge 
any more, and $\kappa(\mu^2) = 0.0854(6)$. This can be seen in fig.~\ref{fig:ghost0s}, where 
the log-log plot of $F_R(0)$ in terms of $\alpha_{\rm crit}-\alpha$ is shown and 
the linear behaviour with negative slope corresponding to the best correlation coefficient
strikingly indicates a zero-momentum ghost propagator diverging 
as $\alpha \to \alpha_{\rm crit}$. Nevertheless, no critical or scaling solution 
appears for the coupled DSE system in the PT-BFM, although the decoupling solutions obtained for any 
$\alpha < \alpha_{\rm crit} = 0.1822$ seem to approach the behaviour of a scaling one 
when $\alpha \to \alpha_{\rm crit}$. This is well understood in ref.~\cite{RodriguezQuintero:2010xx},  
where the gluon propagators obtained 
from the coupled DSE system in PT-BFM were also found to obey the 
same critical behaviour pattern as the ghost propagator, 
when approaching the critical value of the coupling.  .

\section{Conclusions}\label{conclu}

The ghost propagator DSE, with the only assumption of taking $H_1(q,k)$ from the 
ghost-gluon vertex in \eq{DefH12} to be constant in the infrared domain of $q$, can be 
exploited to look into the low-momentum behaviour of the ghost propagator.  The
two classes of solutions named ``decoupling'' and ``scaling'' can be indentified and 
shown to depend on whether the ghost dressing function achieves a finite non-zero
constant ($\alpha_F=0$) at vanishing momentum or not ($\alpha_F \neq 0$). The 
solutions appear to be dialed by the size of the coupling at the renormalization 
momentum which plays the role of a boundary condition for the DSE integration. 
When applying a model with a massive gluon propagator, the decoupling 
low-momentum behaviour of the ghost propagator results to be regulated by the 
gluon mass and the Taylor-scheme effective charge at zero momentum and 
successfully describes the low-momentum ghost propagator computed trhough 
the numerical integration of the coupled gluon and ghost DSE in the PT-BFM scheme. 
The zero-momentum ghost dressing function is then shown to 
tends to diverge when the value of the coupling dialing the solutions 
approaches some critical value. Such a divergent behaviour at the critical coupling 
corresponds to a scaling solution where, if the gluon 
is massive, $\alpha_F=-1/2$.


\begin{theacknowledgments}

The author acknowledges the Spanish MICINN for the 
support by the research project FPA2009-10773 and ``Junta de Andalucia'' by P07FQM02962.

\end{theacknowledgments}



\bibliographystyle{aipproc}   


\begin{thebibliography}{9}

\bibitem{Boucaud:2008ji}
  Ph.~Boucaud, J.~P.~Leroy, A.~L.~Yaouanc, J.~Micheli, O.~Pene and J.~Rodriguez--Quintero,
  JHEP {\bf 0806} (2008) 012
   arXiv:0801.2721 [hep-ph].

\bibitem{Boucaud:2008ky}
  Ph.~Boucaud, J.~P.~Leroy, A.~Le Yaouanc, J.~Micheli, O.~Pene and J.~Rodriguez-Quintero,
  JHEP {\bf 0806} (2008) 099
  [arXiv:0803.2161 [hep-ph]].
  
\bibitem{Aguilar:2006gr}
  A.~C.~Aguilar and J.~Papavassiliou,
  JHEP {\bf 0612} (2006) 012;
  Eur.\ Phys.\ J.\  A {\bf 31} (2007) 742;
A.~C.~Aguilar and A.~A.~Natale,
  JHEP {\bf 0408} (2004) 057.

\bibitem{Aguilar:2008xm}
  A.~C.~Aguilar, D.~Binosi and J.~Papavassiliou,
  Phys.\ Rev.\  D {\bf 78} (2008) 025010
  [arXiv:0802.1870 [hep-ph]].

\bibitem{Alkofer:2000wg}
  R.~Alkofer and L.~von Smekal,
  Phys.\ Rept.\  {\bf 353} (2001) 281
  [arXiv:hep-ph/0007355];
  C.~Lerche and L.~von Smekal, 
  Phys.\ Rev.\ D {\bf 65} (2002) 125006 [arXiv:hep-ph/0202194];
 D.~Zwanziger, 
 Phys.\ Rev.\ D {\bf 65} (2002) 094039 [arXiv:hep-th/0109224];
  C.~S.~Fischer and R.~Alkofer,
  Phys.\ Lett.\  B {\bf 536} (2002) 177
  [arXiv:hep-ph/0202202];
  J.~M.~Pawlowski, D.~F.~Litim, S.~Nedelko and L.~von Smekal,
  Phys.\ Rev.\ Lett.\ {\bf 93} (2004) 152002 [arXiv:hep-th/0312324].
  M.~Q.~Huber, R.~Alkofer, C.~S.~Fischer and K.~Schwenzer,
  Phys.\ Lett.\  B {\bf 659} (2008) 434
  [arXiv:0705.3809 [hep-ph]].

\bibitem{Fischer:2008uz}
  C.~S.~Fischer, A.~Maas and J.~M.~Pawlowski,
  Annals Phys.\  {\bf 324} (2009) 2408
  [arXiv:0810.1987 [hep-ph]].

\bibitem{Cucchieri:2007md}
  A.~Cucchieri and T.~Mendes,
  PoS {\bf LAT2007} (2007) 297
  Phys.\ Rev.\ Lett.\  {\bf 100} (2008) 241601;
%
  I.~L.~Bogolubsky, E.~M.~Ilgenfritz, M.~Muller-Preussker and A.~Sternbeck,
  Phys.\ Lett.\  B {\bf 676} (2009) 69;
  I.~L.~Bogolubsky, E.~M.~Ilgenfritz, M.~Muller-Preussker and A.~Sternbeck,
  PoS {\bf LAT2007} (2007) 290;
%
 A.~Sternbeck, E.-M.~Ilgenfritz, M.~M\:uller-Preussker and A.~Schiller,
Nucl.\ Phys.\ Proc.\ Suppl.\  {\bf 140} (2005) 653;
  AIP Conference Proceedings {\bf 756} (2005) 284,
  [arXiv:hep-lat/0412011];
%
  P.~Boucaud {\it et al.},
 [arXiv:hep-ph/0507104 ];
%
  O.~Oliveira and P.~Bicudo,
  arXiv:1002.4151 [hep-lat];
%
  V.~G.~Bornyakov, V.~K.~Mitrjushkin and M.~Muller-Preussker,
  Phys.\ Rev.\  D {\bf 81} (2010) 054503

\bibitem{Cornwall}
  J.~M.~Cornwall, Phys. Rev. D {\bf 26}, 1453 (1982).




\bibitem{Dudal:2007cw}
  D.~Dudal, J.~A.~Gracey, S.~P.~Sorella, N.~Vandersickel and H.~Verschelde,
  Phys.\ Rev.\  D {\bf 78} (2008) 065047
  [arXiv:0806.4348 [hep-th]].

\bibitem{Frasca:2007uz}
  M.~Frasca,
  Phys.\ Lett.\  B {\bf 670} (2008) 73

\bibitem{Tissier:2010ts}
  M.~Tissier and N.~Wschebor,
  arXiv:1004.1607 [hep-ph].



\bibitem{Boucaud:2010gr}
  Ph.~Boucaud {\it al.}, 
Phys.\ Rev.\  D {\bf 82} (2010) 054007
  [arXiv:1004.4135 [hep-ph]]

\bibitem{RodriguezQuintero:2010xx}
J.~Rodriguez-Quintero, 
[arXiv:1005.4598 [hep-ph]].

\bibitem{Binosi:2002ft}
  D.~Binosi and J.~Papavassiliou,
  Phys.\ Rev.\  D {\bf 66} (2002) 111901
  [arXiv:hep-ph/0208189];
  Phys.\ Rev.\  D {\bf 77} (2008) 061702
  [arXiv:0712.2707 [hep-ph]];
  D.~Binosi and J.~Papavassiliou,
  Phys.\ Rev.\  D {\bf 77} (2008) 061702
  [arXiv:0712.2707 [hep-ph]].



\bibitem{Boucaud:2008gn}
  Ph.~Boucaud {\it et al.}, 
  Phys.\ Rev.\  D {\bf 79} (2009) 014508
  [arXiv:0811.2059 [hep-ph]];
  A.~Sternbeck, K.~Maltman, L.~von Smekal, A.~G.~Williams, E.~M.~Ilgenfritz and M.~Muller-Preussker,
  PoS {\bf LAT2007} (2007) 256
  [arXiv:0710.2965 [hep-lat]].

\bibitem{Aguilar:2008fh}
  A.~C.~Aguilar, D.~Binosi and J.~Papavassiliou,
  PoS {\bf LC2008} (2008) 050
  [arXiv:0810.2333 [hep-ph]].

\bibitem{Aguilar:2009nf}
  A.~C.~Aguilar, D.~Binosi, J.~Papavassiliou and J.~Rodriguez-Quintero,
  Phys.\ Rev.\  D {\bf 80} (2009) 085018
    [arXiv:0906.2633 [hep-ph]].




\end{thebibliography}

\IfFileExists{\jobname.bbl}{}
 {\typeout{}
  \typeout{******************************************}
  \typeout{** Please run "bibtex \jobname" to optain}
  \typeout{** the bibliography and then re-run LaTeX}
  \typeout{** twice to fix the references!}
  \typeout{******************************************}
  \typeout{}
 }


\end{document}